\documentstyle[preprint,aps]{revtex}
\input prepictex
\input pictex
\input postpictex
\begin{document}
\title{Hole dynamics in generalized spin backgrounds in infinite dimensions}

\author{Karen A. Hallberg}
\address{Max-Planck-Institut f\"{u}r Physik komplexer Systeme,
Bayreuther Strasse 40,
Haus 16, 01187 Dresden, Germany.}
\author{Erwin M\"uller-Hartmann}
\address{Institut f\"ur Theoretische Physik, Universit\"at zu K\"oln,
Z\"ulpicher Strasse 77, 50937 K\"oln}
\author{C. A. Balseiro}
\address{Centro At\'{o}mico Bariloche and Instituto Balseiro, 8400 San
Carlos de Bariloche, Argentina.}
\date{\today}
\maketitle

\begin{abstract}

We calculate the dynamical behaviour of a hole in various spin backgrounds
in infinite dimensions, where it can be determined exactly.
 We consider hypercubic lattices with
two different types of spin backgrounds. On one hand we study an
ensemble of spin configurations with an arbitrary spin probability on each
sublattice. This model
 corresponds to a thermal average over all spin
configurations in the presence of  staggered or uniform
magnetic fields. On the other hand we
consider a definite spin state characterized by the angle between the spins
on different sublattices, {\it i.e} a classical spin system
in an external magnetic field.
 When spin fluctuations are considered,
this  model describes the physics of unpaired particles in
strong coupling superconductors.

\end{abstract}
\pacs{75.10.Jm, 71.10.+x}

\narrowtext

\section{Introduction}
\label{sec:Introduction}

	It is well known that the discovery of high Tc superconductors has
triggered an extensive study of highly correlated systems. The Hubbard and
$t-J$ models have been prototype models towards the
understanding of most of the features of
those materials. For low doping (near half filling) the system consists of
dilute mobile holes in a spin background.

A lot of effort has been devoted to the understanding of the dynamics of
holes in spin backgrounds. Brinkman and Rice (BR) \cite{br} considered several
configurations of the spin background (ferromagnetic (FM), N\'eel  and
random) and studied the density of hole-states and dc conductivity. Their
calculation was based on Nagaoka's expansion of expectation values in terms
of background-conserving hole paths\cite{naga}. Of these, they only considered
retraceable paths, {\it i.e.}, no loops were taken into account. They obtained
very accurate results for single-particle Green's functions of a hole in the
N\'eel background, being exact in one dimension (where all paths are
retraceable). This retraceable paths approximation (rpa) was used to study
many other quantities such as dynamical conductivity \cite{rice},
electrical resistivity, thermal conductivity, thermopower and
specific heat \cite{oguri}.

Using different approximations, many other analytical approaches have been
reported for the ground state and excited properties of a single hole
\cite{6y,7y,11y}. There are also numerous studies using exact diagonalization
techniques in low dimensions \cite{8y,nuestro,16mv,17mv}. These results
show a well defined quasiparticle peak for $J\simeq t$ whereas for small
enough $J/t$ an incoherent spectrum carries most of the spectral weight.

	Most of the studies have been concentrated on the dynamics of a hole
in an antiferromagnetic (AF) background. The case of the polarized $t-J$ model
has been studied in connection with strong coupling superconductors
\cite{nuestro,12y}. The negative $U$ Hubbard model with $n$ particles is
equivalent to a positive-$U$ case with one particle per site and a net
magnetization given by $S_z={(1-n)\over 2}$ \cite{13y}. So the dynamics of
a single unpaired particle moving in a background of strongly bounded paired
particles (strong negative $U$ limit) is described by the $t-J$ model.

	In the present work we study the single-particle Green's functions
in certain backgrounds, extending the results of Metzner et al.
\cite{metzner} to generalized spin backgrounds and in particular to one that
describes the physics of holes in strong coupling superconductors.

 The calculations are performed for infinite dimensions
where the
results are exact. The limit of high lattice dimensions, $d\to \infty$, has
been used to study correlated fermions \cite{21mv} and helped to clarify the
validity of several approximations and construct new ones \cite{26mv}.

A self-consistent approximation for finite dimensions can be performed in
a similar way as in Ref. \cite{metzner}.
More realistic calculations applicable to high Tc superconductors,
for example, should also include interactions between holes and spin
fluctuations. We neglect spin fluctuations since
they disappear in infinite dimensions, and concentrate on the corrections
due to the inclusion of loops to the BR retraceable path approximation.
This latter approach doesn't distinguish between different spin backgrounds
since paths without loops are always background conserving. While in the
N\'eel background the BR approximation is correct up to order $1/d^4$, where
$d$ is the dimension, the contribution of loops becomes important
whenever there are clusters of aligned spins. In particular, in a FM
background,
any hole path leaves the background unchanged, leading to Nagaoka's theorem
\cite{naga}.

	The paper is organized as follows: in Sec.~II we calculate the dynamics
of a hole in an ensemble of spin configurations in a hypercubic lattice,
considering arbitrary spin averages or probabilities
in each sublattice; in Sec.~III we consider a definite spin state
characterized by the
angle between the spins on different sublattices and we obtain expressions
for the local and ${\bf k}-$dependent propagator. We summarize in Sec.~IV.

\section{One hole in a generalized spin background}

	We consider the $t-J$ model for infinite dimensions with $J=0$, which
with standard notation reads:
\begin{equation}
H=-t\sum_{<ij>,\sigma}(1-n_{i-\sigma})c^\dagger_{i\sigma}c_{j\sigma}(
1-n_{j-\sigma})
\end{equation}

	To keep the average kinetic energy finite in $d\to\infty$, one must
scale the hopping amplitude $t$ as
\begin{equation}
\label{eq:t}
t={t^*\over \sqrt{Z}}
\end{equation}
$t^*$ fixed, where $Z$ is the number of nearest neighbours ($Z=2d$ in a
hypercubic lattice) \cite{21mv}.

	The Green's function of a hole in an arbitrary spin background S is:
\begin{eqnarray}
G_{ij}^S(z)&=&\sum_{\sigma}G_{ij \sigma}^S(z) \nonumber \\
G_{ij \sigma}^S(z)&=&\langle c^{\dagger}_{i\sigma}{1\over z-H}c_{j\sigma}
\rangle_S
\end{eqnarray}
where $\langle \cdots \rangle_S=\sum_{S'}w_{S'}\langle S'|\cdots|S'\rangle$
and $w_{S'}$ is a normalized distribution of spin configurations $S'$. The
spins of different sites are statistically independent. We consider a
hypercubic lattice and characterize the spin ensembles by asigning a
probability for spin $\sigma$ in the sublattice $X$: $p_{X\sigma}$;
$X=(A,B)$ and $\sigma=(\uparrow,\downarrow)$. The random case considered
in Ref.\cite{metzner} corresponds to $p_{A\sigma}=p_{B\sigma}$ and interpolates
between the FM ($p_{A\sigma}=1$) and the unpolarized random ($p_{A\sigma}=
1/2$). We generalize the calculation to interpolate also between the random
and the N\'eel
background ($p_{A\sigma}=p_{B-\sigma}=1$). The unpolarized random case
corresponds to a thermal average over all spin configurations which contribute
with equal weight. The other polarized cases correspond to thermal averages
in presence of uniform or staggered magnetic fields.

We follow closely Ref.\cite{metzner} for the notations. The calculations are
based principally on Nagaoka's expansion for the Green's functions in powers
of $t/z$ \cite{naga}. For our hypercubic lattice we have to distinguish two
local Green's functions:
\begin{equation}
\label{eq:gaa}
G_{AA}(z)={1\over z[1-S_A(z)]} \;\;\; \mbox{and}\;\;\;
G_{BB}(z)={1\over z[1-S_B(z)]}
\end{equation}
where $S_X \;\;(X=A,B)$ is given by the sum over all graphs for which the hole
returns to the starting point only once. Due to the simple topology of the
loop trees, $S_X$ can be written as a sum over loops with dressed vertices
$C_X$. This dressed vertex is given by the bare one plus all possible $S_X$
insertions, so we have:
\begin{equation}
\label{eq:cx}
C_{X\sigma}(z)=p_{X\sigma}\sum_{l=0}^\infty [S_X(z)]^l=
{p_{X\sigma}\over 1-S_X(z)}
\end{equation}
and
\begin{mathletters}
\label{eq:sa}
\begin{eqnarray}
S_A(z)&=&\sum_\sigma\sum_{n=1}^\infty u_{2n}[C_{B\sigma}^n C_{A\sigma}^{n-1}
]\left ({t^*\over z}\right)^{2n} \\
S_B(z)&=&\sum_\sigma\sum_{n=1}^\infty u_{2n}[C_{A\sigma}^n C_{B\sigma}^{n-1}
]\left ({t^*\over z}\right)^{2n}
\end{eqnarray}
\end{mathletters}
where $u_n$ is the number of self-avoiding return paths of length $n$.

Defining a generating function for $u_n$
\begin{equation}
\label{eq:m}
M(\xi)=1+\sum_{n=2}^\infty u_n \xi^n
\end{equation}
equation (\ref{eq:sa}) can be written as
\begin{mathletters}
\label{eq:sa2}
\begin{eqnarray}
S_A(z)&=&\sum_\sigma {M[\frac{t^*}{z}\sqrt{C_{A\sigma}C_{B\sigma}}]-1\over
C_{A\sigma}} \\
S_B(z)&=&\sum_\sigma {M[\frac{t^*}{z}\sqrt{C_{A\sigma}C_{B\sigma}}]-1\over
C_{B\sigma}}
\end{eqnarray}
\end{mathletters}

{}From Eqs.~(\ref{eq:gaa}), (\ref{eq:cx}) and (\ref{eq:sa2}) and
defining $G^0$ as the free particle Green's function and $G^0_{-1}$ its
inverse, the following relations hold
in the FM limit ($p_{A\downarrow}=p_{B\downarrow}\to 0$)
\begin{equation}
M(\xi)=1+{\cal O}(\xi^2) \;\;\mbox{for small $\xi$}\;\; \Rightarrow
M(tG^0)=zG^0
\end{equation}
This implies
\begin{equation}
\label{eq:g0}
M(\xi)={\xi \over t^*}G^0_{-1}(\xi/t^*)
\end{equation}

Using this relation in a similar way as in  \cite{metzner} we find:
\begin{mathletters}
\label{eq:pau}
\begin{eqnarray}
1+p_{A\uparrow}p_{A\downarrow}(zG_{AA}-1)&=&\sum_\sigma p_{A-\sigma}
\sqrt{p_{A\sigma}p_{B\sigma}G_{AA}G_{BB}}G^0_{-1}\left[
\sqrt{p_{A\sigma}p_{B\sigma}G_{AA}G_{BB}}\right]  \\
1+p_{B\uparrow}p_{B\downarrow}(zG_{BB}-1)&=&\sum_\sigma p_{B-\sigma}
\sqrt{p_{A\sigma}p_{B\sigma}G_{AA}G_{BB}}G^0_{-1}\left[
\sqrt{p_{A\sigma}p_{B\sigma}G_{AA}G_{BB}}\right]
\end{eqnarray}
\end{mathletters}

	A more compact expression can be obtained in the symmetric case
where $p_{A\sigma}=p_{B-\sigma}$. This implies $G_{AA}=G_{BB}=G$ and
Eqs.~(\ref{eq:pau}) simplify to
\begin{equation}
\label{eq:gamma}
\gamma G=G^0\left[ {1+\gamma^2(zG-1)\over \gamma G}\right]
\end{equation}
with $\gamma=\sqrt{p_{A\sigma}p_{A-\sigma}}$ ranging from N\'eel to
uniform random backgrounds ($0\le \gamma \le 1/2$).
The density of states
$D(\omega)=-{1 \over \pi}Im G(\omega+i0^+)$ for this symmetric case
coincides exactly with the density of states of the tilted configuration
with the correspondence $\gamma=\alpha$ for $\alpha\le 1/2$
(see Eq.~(\ref{eq:alphagamma}) and Fig.~2).

As for $|\omega|\gg t^*\Rightarrow
G^0(\omega)\sim 1/\omega $, Eq.~(\ref{eq:gamma}) implies $G(\omega)\sim
1/\omega$ and $\gamma G\sim G^0(\omega/\gamma)$. Using the density of states of
a free particle at large $|\omega |$ we have:
\begin{equation}
D(\omega)\sim {1\over \sqrt{2\pi}\gamma t^*}e^{-\omega^2/(2\gamma^2 t^{*2})}
\end{equation}
The density of states has exponential tails for large $|\omega|$ whenever
$\gamma\neq 0$. These come from the presence of FM clusters in the spin
configuration.

In the N\'eel limit ($p_{A\downarrow}\to 0 \Rightarrow \gamma\to 0$),
to first order in $p_{A\downarrow}$, Eq.~(\ref{eq:pau}) leads to
\begin{equation}
\label{eq:neel}
zG-1=t^{*2} G^2 \Rightarrow G={z-\sqrt{z^2-4t^{*2}}\over 2t^{*2}}
\end{equation}
This coincides with the results of the rpa \cite{br}.
The corresponding density of states
\begin{equation}
D(\omega)={1\over 2\pi t^{*2}}\sqrt{4t^{*2}-\omega^2}; \;\;\;|\omega|
\le 2t^*
\end{equation}
has a semielliptic shape with band edges at $\pm 2t^*$ with a square root
singularity.

We also calculated the self-energy for this symmetric case. It is
$k$-independent for \ $d\to\infty$ and from symmetry $\Sigma_{A\sigma}=
\Sigma_{B-\sigma}$. It is not necessary to use the Nagaoka expansion for the
off-diagonal propagator in the site representation as we did for the
calculation of the local propagator
\cite{metzner}. Instead,
we use the Dyson equation together with the locality of the self energy.

In a matrix representation for the $k$-dependent Green's function
in the reduced Brillouin zone
( $\epsilon_k=-2t(\cos k_1+\cdots+\cos k_d.< 0$) we have
\begin{equation}
G_{{\bf k}\sigma}^{-1}(z)=\left (\begin{array}{cc}
                           z-\Sigma_{A\sigma} & \epsilon_k \\
                           \epsilon_k         & z-\Sigma_{B\sigma}
                            \end{array} \right )
\end{equation}
This leads to
\begin{equation}
G_{{\bf k}\sigma}(z)={1\over D}\left (\begin{array}{cc}
                           z-\Sigma_{B\sigma} & -\epsilon_k \\
                           -\epsilon_k         & z-\Sigma_{A\sigma}
                            \end{array} \right )
\end{equation}
\begin{equation}
D=\left (\sqrt{(z-\Sigma_{A\sigma})(z-\Sigma_{B\sigma})}-\epsilon_k\right )
  \left (\sqrt{(z-\Sigma_{A\sigma})(z-\Sigma_{B\sigma})}+\epsilon_k\right )
\end{equation}
After some algebra and considering the fact that
\begin{eqnarray}
\label{eq:g02}
\langle {1\over a+\epsilon_k}\rangle _k^-=
\int_{-\infty}^0 {D^0(\epsilon)\over a+\epsilon}=
\int_0^\infty {D^0(\epsilon)\over a-\epsilon}=
\langle {1\over a-\epsilon_k}\rangle _k^+  \nonumber \\
\langle {1\over a-\epsilon_k} + {1\over a+\epsilon_k} \rangle _k^-=
\langle {1\over a-\epsilon_k}\rangle _k=G^0(a)
\end{eqnarray}
and that $G_{X\sigma}=\langle G_{{\bf k}\sigma} \rangle_k=
 p_{X\sigma} G$ we obtain:
\begin{mathletters}
\label{eq:pag}
\begin{eqnarray}
\sqrt{{z-\Sigma_{B\sigma}\over z-\Sigma_{A\sigma}}}G^0\left [
\sqrt{(z-\Sigma_{A\sigma})(z-\Sigma_{B\sigma})} \right ]&=&p_{A\sigma} G \\
\sqrt{{z-\Sigma_{A\sigma}\over z-\Sigma_{B\sigma}}}G^0\left [
\sqrt{(z-\Sigma_{A\sigma})(z-\Sigma_{B\sigma})} \right ]&=&p_{B\sigma} G
\end{eqnarray}
\end{mathletters}
{}From Eqs.~(\ref{eq:gamma}) and (\ref{eq:pag}) we obtain:
\begin{mathletters}
\begin{eqnarray}
\Sigma_{A\sigma}=z(1-p_{B\sigma})-{1-\gamma^2\over p_{A\sigma}G} \\
\Sigma_{B\sigma}=z(1-p_{A\sigma})-{1-\gamma^2\over p_{B\sigma}G}
\end{eqnarray}
\end{mathletters}

These expressions lead to the known results for the N\'eel limit ($\gamma
=0$) and the random case ($\gamma=1/2$) (see Ref.\cite{metzner}). Here
$Im \Sigma_{X\sigma}(\omega-i0^+)>0 \;\forall \omega $ and there are no
quasiparticles in the system. This is expected since this system doesn't
have a Fermi surface.

\section{One hole in a tilted spin background}

	A more compact and straightforward calculation can be done by
considering a magnetic field acting on the system polarizing the spins
in a way sketched in Fig.~1 . As in the previous section, we are considering
a hypercubic lattice in a pure spin configuration
with spins $\sigma_A$ and $\sigma_B$ lying on the $yz$ plane, forming an angle
$\phi$ betwen them. In this way we can continuously connect
the N\'eel ($\phi=\pi$) to the
FM ($\phi=0$) background.

As we mentioned in the Introduction,
this model also describes the dynamics
of an unpaired particle in a strong coupling superconductor. The spin
variables up and down correspond to an empty and a doubly occupied site,
respectively. The external magnetic field plays the role of a chemical
potential since the total magnetization is given by the total number of
particles $S_z=(1-n)/2$. The magnetization in the $x-y$ plane
corresponds to the superfluid order parameter.

By moving along a loop, a hole will replace $A$-spins with $B$-spins and
the overlap between two states at any given site is
\begin{equation}
\alpha=\langle\uparrow| e^{i\phi\sigma_x/2}|\uparrow\rangle=
\cos({\phi\over 2})
\end{equation}

For a $2n$-loop, $2n-2$ sites are changed, therefore
\begin{equation}
\label{eq:s}
S(z)=\sum_{n=1}^\infty u_{2n}C^{2n-1}\alpha^{2n-2}\left
({t^*\over z}\right )^{2n}={M\left [\alpha Ct^*/z \right ]-1 \over
\alpha^2 C}
\end{equation}
where $M$ is given by Eq.~(\ref{eq:m}).

Here again
\begin{equation}
\label{eq:g}
G(z)={1\over z(1-S(z))} \;\;\; \mbox{and}\;\;\; C(z)={1\over 1-S(z)}
\end{equation}
{}From Eqs.~(\ref{eq:g0}), (\ref{eq:s}) and (\ref{eq:g}) we obtain
\begin{equation}
\label{eq:alpha}
\alpha G(z)=G^0\left [ \alpha z + {1-\alpha^2\over \alpha G(z)} \right ]
\end{equation}

This expression is equal to Eq.~(\ref{eq:gamma}) for the symmetric random
model with the correspondence
\begin{equation}
\label{eq:alphagamma}
\alpha=\cos(\phi /2)=\gamma = \sqrt{p_{A\sigma}p_{A-\sigma}}
\end{equation}
for $\alpha\le 1/2$. But Eq.~(\ref{eq:alpha}) is more general because
$\alpha$ can assume any value between $0$ and $1$:
$\alpha=1$
corresponds to the FM case and $\alpha=0$ to the N\'eel case. The hole dynamics
for $\alpha=1/2$ $(\Rightarrow \phi=2\pi/3)$ happens to coincide exactly
with that for the unpolarized random background.

In Fig.~2 we show the density of states corresponding to the one-particle
Green function of Eq.~(\ref{eq:alpha}) for several values of $\phi$. $G$ has
been calculated self-consistently using the N\'eel Green function
(\ref{eq:neel}) as a seed. The only case without tails is the N\'eel
configuration ($\phi=\pi$) that presents square root singularities at the
band edges. The states in the tails of the other configurations correspond
to low energy (say $\omega\to -\infty$) states due to FM clusters. It also
seems that for all values of $\alpha$ there is an energy $\omega_0$ such that
$D(\omega_0)$ is independent of $\alpha$ (this crossing
also happens for the real part of the propagator at another energy).
But this coincidence is only
approximate, being, nevertheless, a curious feature.

We calculated the full propagator in the tilted background. For the
disposition of axes chosen, the spin on the $A$-sites can be written as:
\begin{equation}
e^{i\sigma_x\phi/4}|\uparrow\rangle=\cos({\phi\over 4})|\uparrow\rangle
-\sin({\phi\over4})|\downarrow\rangle
\end{equation}
and on the $B$-sites by changing $\phi$ by $-\phi$. So the Green function
can be written in a matrix way in the spin representation
\begin{mathletters}
\label{eq:gss}
\begin{equation}
G_{AA\sigma\sigma'}=G \left(\begin{array}{cc}
       \cos^2({\phi\over 4}) & -\sin({\phi\over 4})\cos({\phi\over 4}) \\
      -\sin({\phi\over 4})\cos({\phi\over 4}) & \sin^2({\phi\over 4})
       \end{array} \right)
\end{equation}
\begin{equation}
G_{BB\sigma\sigma'}=G (\phi\leftrightarrow -\phi)
\end{equation}
\end{mathletters}
For this type of configurations there is a generalized Bloch theorem. The
elements of the symmetry group that leave the configuration $|X\rangle$ and
the Hamiltonian invariant are $g_{\bf l}=T_{\bf l} e^{il\pi\sigma_z/2}$,
where $ l=\sum_{n=1}^d l_n$; {\it i.e.} translations coupled with
rotations in $\pi$ about the $z$ axis. So $c^\dagger_{{\bf k}\sigma}=
{1\over \sqrt{V}}\sum_{\bf l}c^\dagger_{{\bf l}\sigma}e^{i{\bf k}\cdot{\bf l}}$
 is an eigenoperator of $G_{\bf l}$ with eigenvalue ${\bf q}$ such that
$e^{i{\bf q}\cdot{\bf l}}=(\sigma i)^l e^{i{\bf k}\cdot{\bf l}}=
e^{i({\bf k}+\sigma {\bf Q}/2)\cdot {\bf l}}$ and ${\bf Q}=(\pi,\cdots \pi)$.
This implies that $c^\dagger_{{\bf k}\sigma}$ can couple only to
$c^\dagger_{{\bf k+Q}-\sigma}$.

We define a propagator in the spin representation
\begin{equation}
\label{eq:gs}
G_{{\bf k}\sigma\sigma'}=\left (\begin{array}{cc}
\langle\langle c^\dagger_{{\bf k}\uparrow}|c_{{\bf k}\uparrow}\rangle\rangle
& \langle\langle c^\dagger_{{\bf k}\uparrow}|
c_{{\bf k+Q}\downarrow}\rangle\rangle \\
\langle\langle c^\dagger_{{\bf k+Q}\downarrow}|c_{{\bf k}\uparrow}
\rangle\rangle
&\langle\langle c^\dagger_{{\bf k+Q}\downarrow}|
c_{{\bf k+Q}\downarrow}\rangle\rangle
\end{array} \right )
\end{equation}
Then
\begin{equation}
\label{eq:<>k}
\langle G_{{\bf k}\sigma\sigma'}\rangle_k=G_{AA\sigma\sigma'}=
\sigma \sigma' G_{BB\sigma\sigma'}
\end{equation}
Considering the locality of the self energy we can write the inverse of Eq.~
(\ref{eq:gs}) as:
\begin{equation}
G_{{\bf k}\sigma\sigma'}^{-1}(z)=\left ( \begin{array}{cc}
z-\epsilon_k-\Sigma_{\uparrow\uparrow} & -\Sigma_{\uparrow\downarrow} \\
-\Sigma_{\downarrow\uparrow} &  z+\epsilon_k-\Sigma_{\downarrow\downarrow}
\end{array} \right )
\end{equation}
so
\begin{equation}
G_{{\bf k}\sigma\sigma'}(z)={1\over D}\left ( \begin{array}{cc}
z+\epsilon_k-\Sigma_{\downarrow\downarrow} & \Sigma_{\downarrow\uparrow} \\
\Sigma_{\uparrow\downarrow} &  z-\epsilon_k-\Sigma_{\uparrow\uparrow}
\end{array} \right )
\end{equation}
\begin{equation}
D=(z-\epsilon_k-\Sigma_{\uparrow\uparrow})(z+\epsilon_k-\Sigma_
{\downarrow\downarrow})-\Sigma_{\uparrow\downarrow}\Sigma_{\downarrow
\uparrow}=-(\epsilon_k-E_+)(\epsilon_k-E_-)
\end{equation}
where
\begin{equation}
\label{eq:e}
E_{\pm}={1\over 2}(\Sigma_{\downarrow\downarrow}-\Sigma_{\uparrow\uparrow})
\pm {1\over 2}\sqrt{(\Sigma_{\downarrow\downarrow}-\Sigma_{\uparrow\uparrow}
)^2-4\Sigma_{\uparrow\downarrow}\Sigma_{\downarrow\uparrow}+
4(z-\Sigma_{\uparrow\uparrow})(z-\Sigma_{\downarrow\downarrow})}
\end{equation}
This leads to
\begin{equation}
G_{{\bf k}\sigma\sigma'}={1\over E_+-E_-}\left (\begin{array}{cc}
-{z+E_--\Sigma_{\downarrow\downarrow}\over E_--\epsilon_k}+
{z+E_+-\Sigma_{\downarrow\downarrow}\over E_+-\epsilon_k}&
\;\Sigma_{\downarrow\uparrow}\left ( -{1\over E_--\epsilon_k}+
{1\over E_+-\epsilon_k} \right ) \\
\Sigma_{\uparrow\downarrow}\left ( -{1\over E_--\epsilon_k}+
{1\over E_+-\epsilon_k} \right ) &
\;-{z-E_--\Sigma_{\uparrow\uparrow}\over E_--\epsilon_k}+
{z-E_+-\Sigma_{\uparrow\uparrow}\over E_+-\epsilon_k}
\end{array} \right )
\end{equation}

Using Eqs.~(\ref{eq:g02}), (\ref{eq:<>k}) and (\ref{eq:e}) we obtain
\begin{equation}
\label{eq:gaa2}
G_{AA\sigma\sigma'}=-{1\over \sqrt{W}}\left (\begin{array}{cc}
{\scriptstyle
\left [z-{1\over 2}(\Sigma_{\uparrow\uparrow}+\Sigma_{\downarrow\downarrow})
\right ](G^0_--G^0_+)-{1\over 2}\sqrt{W}(G^0_-+G^0_+)} &
{\scriptstyle \Sigma_{\downarrow\uparrow}(G^0_--G^0_+)} \\
{\scriptstyle \Sigma_{\uparrow\downarrow}(G^0_--G^0_+)} &
{\scriptstyle
\left [z-{1\over 2}(\Sigma_{\uparrow\uparrow}+\Sigma_{\downarrow\downarrow})
\right ](G^0_--G^0_+)+{1\over 2}\sqrt{W}(G^0_-+G^0_+) }
\end{array} \right )
\end{equation}
where
\begin{equation}
W=(\Sigma_{\downarrow\downarrow}-\Sigma_{\uparrow\uparrow})^2-
4\Sigma_{\uparrow\downarrow}\Sigma_{\downarrow\uparrow}+
4(z-\Sigma_{\uparrow\uparrow})(z-\Sigma_{\downarrow\downarrow})
\end{equation}
and we have defined $G^0(E_{\pm})=G^0_{\pm}$.

Eq.~(\ref{eq:gaa2}) implies $\Sigma_{\uparrow\downarrow}(z)=
\Sigma_{\downarrow\uparrow}(z)$.
Together with Eq.~(\ref{eq:gss}) it also gives a set of three equations for
$\Sigma_{\uparrow\uparrow}$, $\Sigma_{\downarrow\downarrow}$ and
$\Sigma_{\uparrow\downarrow}$.
\begin{mathletters}
\label{eq:3}
\begin{equation}
(2z-\Sigma_{\uparrow\uparrow}-\Sigma_{\downarrow\downarrow})
(G^0_+-G^0_-)=G \sqrt{W}
\end{equation}
\begin{equation}
G^0_++G^0_-=G\cos({\phi\over 2})
\end{equation}
\begin{equation}
2\Sigma_{\uparrow\downarrow} (G^0_+-G^0_-)=-G\sin({\phi\over 2})\sqrt{W}
\end{equation}
\end{mathletters}
{}From these equations we obtain $G^0_-=0$ which implies $E_-\to \infty$
as well as the three self energies.
{}From $G^0(E_+)=G\cos({\phi\over 2})$ and Eq.~(\ref{eq:alpha}) it follows that
\begin{equation}
\label{eq:emas}
E_+=z\cos({\phi\over 2})+{\sin^2({\phi\over 2}) \over G\cos({\phi\over 2})}
\end{equation}

We cannot obtain separate expressions for the self energies but
$G_{{\bf k}\sigma\sigma'}$ can be calculated by knowing the ratios
$\Sigma_{\uparrow\uparrow}/E_-$,
$\Sigma_{\downarrow\downarrow}/E_-$ and
$\Sigma_{\uparrow\downarrow}/E_-$. From Eq.~(\ref{eq:3}) and the definition
of $E_-$ (\ref{eq:e}), we have, for $E_-\to \infty$
\begin{mathletters}
\begin{equation}
\Sigma_{\downarrow\downarrow}/E_-\to \left (1-\tan^2({\phi\over 4})\right )
^{-1}
\end{equation}
\begin{equation}
\Sigma_{\uparrow\uparrow}/E_-\to \left (\cot^2({\phi\over 4})-1\right )^{-1}
\end{equation}
\begin{equation}
\Sigma_{\uparrow\downarrow}/E_-\to \left (2\cot({\phi\over 2})\right )^{-1}
\end{equation}
\end{mathletters}

Finally the full propagator becomes:
\begin{equation}
G_{{\bf k}\sigma\sigma'}={\tan({\phi\over 2})\over 2(E_+-\epsilon_k)}
\left ( \begin{array}{cc}
\cot(\phi/4) &
-1 \\
-1 &
\tan(\phi/4)
\end{array} \right )
\end{equation}
where $E_+$ is given by Eq.~({\ref{eq:emas}). From Eq.~(\ref{eq:gs}) we see
that the spin-flipped propagators are obtained by replacing ${\bf k}$ by
${\bf k+Q}$, {\it i.e.} $\epsilon_k$ by $-\epsilon_k$.

In Fig.~3 we show the spectral density $\rho_{\bf k}(\omega)$ obtained from
these equations for several values of $\phi$ and ${\bf k}={\bf Q}$. The
N\'eel case is equal to the local density of states since it is k-independent.
In particular, no quasiparticle peak is present since the system has no Fermi
surface.
 Only the FM background presents a quasiparticle peak that
corresponds to the dynamics of a free particle.

\section{Summary and Conclusions}

	We have obtained general expressions for the single particle Green's
functions describing the dynamics of a hole in a generalized spin
background in infinite dimensions. Our calculations are a generalization
of those performed in Ref.\cite{metzner}.

We characterized the ensemble of spin configurations
in a hypercubic lattice by the
average value of the spin in each sublattice. Closed expressions
were found for the symmetric case ($p_{A\sigma}=p_{B-\sigma}$),
interpolating
between N\'eel order and the unpolarized random case.

We also performed the calculations considering a magnetic field acting on the
system that tilts the spins in such a way that spins of different sublattices
form an angle $\phi$ between them. In this case we obtain expressions for
the local and k-dependent propagator as a function of the angle $\phi$, so the
results interpolate between the FM and N\'eel configurations.

The local density of states gets narrower when departing from the N\'eel
order and acquires exponential tails that come from the contribution of FM
clusters to the hole motion. The spectral function $\rho_{\bf k}(\omega)$
also shows this behaviour. A quasiparticle peak is seen only in the FM case
and it corresponds to the movement of a free particle.

A self-consistent calculation for finite dimensions can also be carried out
in a similar way as in Ref.\cite{metzner}, considering the exact expression
for $G^0$ in $d$-dimensions and scaling the hopping as in Eq.~(\ref{eq:t}).
Nevertheless, one has to take into account that in this calculation
only loop trees have been considered.
Paths with loops on which a
hole walks around more than once or with multiply connected loops, are
suppressed  with respect to the loop tree
\cite{27mv} by some integer power of $1/d$.
Also, in the N\'eel case, at the band edges, where the density
of states is small, corrections become very large. Another important
feature that is neglected in these calculations is the presence of spin
fluctuations that become important at low dimensions. Nevertheless, this
approximation can always be corrected by properly taking into account
all the relevant hole paths (for the 2D N\'eel case see Ref.~\cite{ces})
or by allowing background restoring spin flips along the path.

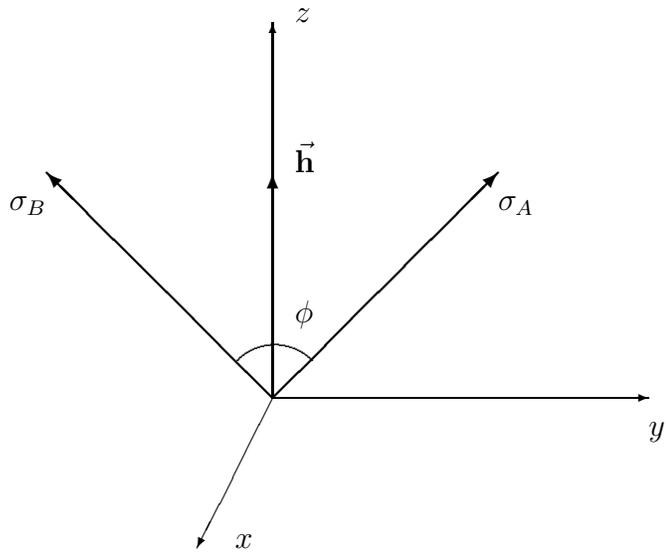
\begin{figure}
\caption{Schematic representation of the spin configuration in
the tilted background, in the presence of a magnetic field $h$.}
%\label{fig:1}
\end{figure}
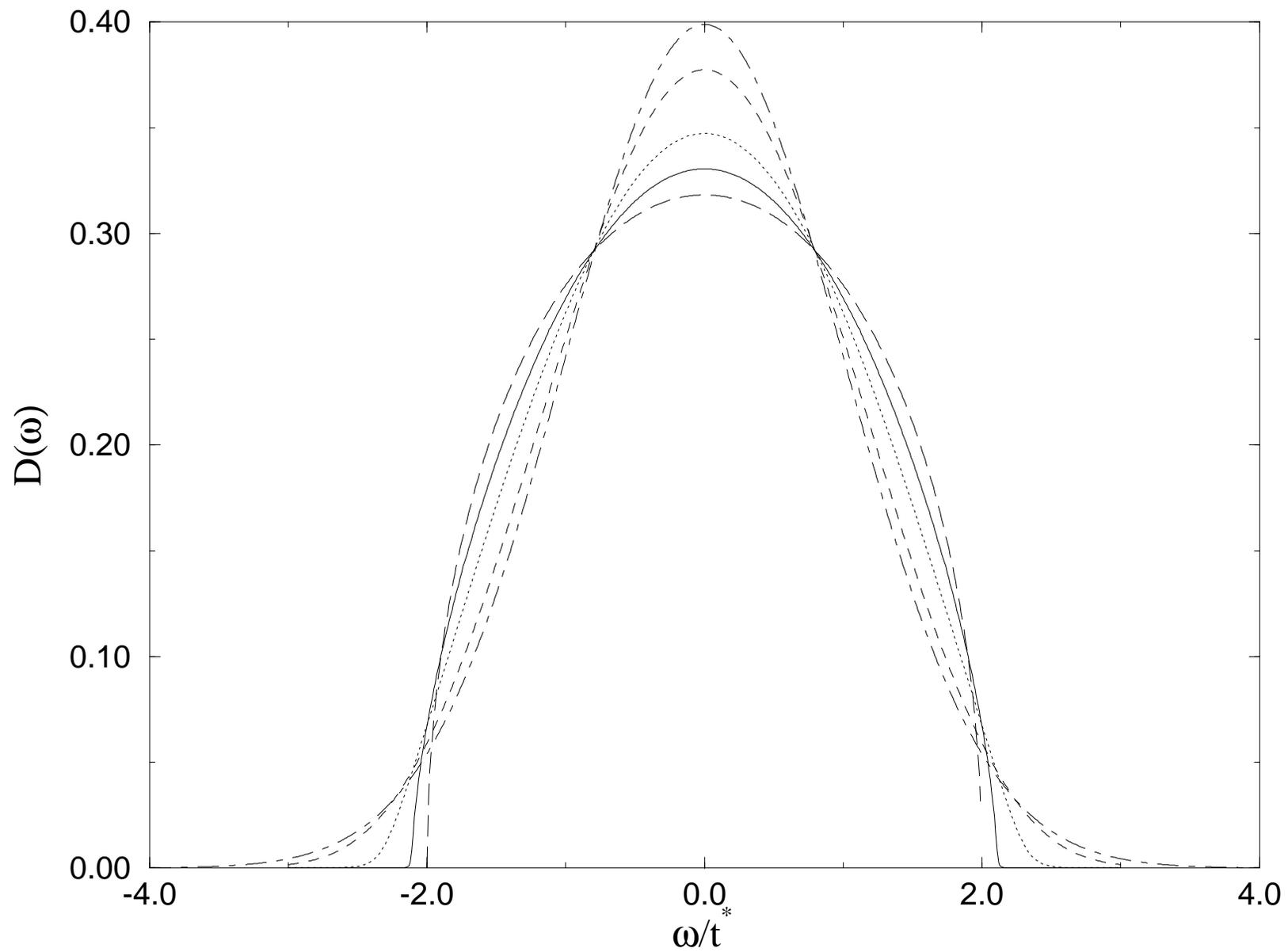
\begin{figure}
\caption{Density of states for the tilted background for different values
of $\alpha=\cos(\phi /2)$: $\alpha=1$ (FM case, dashed-dotted line),
$\alpha=.8$ (dashed), $\alpha=.5$ (dotted), $\alpha=.3$ (full),
$\alpha=0$ (N\'eel, long-dashed line).}
%\label{fig:2}
\end{figure}
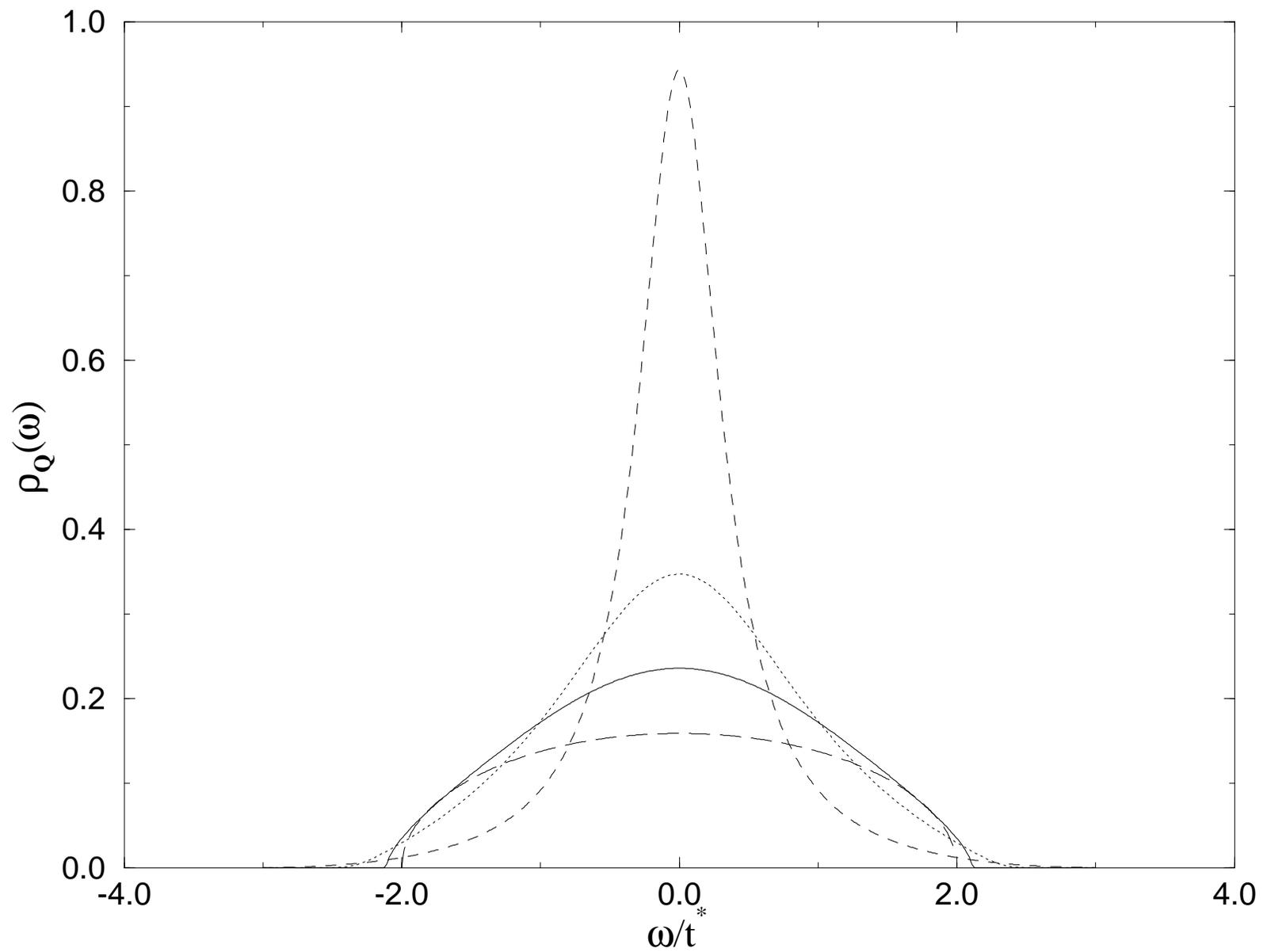
\begin{figure}
\caption{Spectral density for the tilted background for ${\bf k}={\bf Q}$
and different values
of $\alpha=\cos(\phi /2)$: The symbols correspond to those of Fig.~2. The
FM case presents a delta function at $\omega=0$ }
%\label{fig:3}
\end{figure}

\newpage

\setlength{\unitlength}{1mm}
\setcoordinatesystem units <1mm,1mm> point at 50 30
\beginpicture
\put{
\begin{picture}(150,100)(0,0)
\thinlines
\put(50,30){\vector(1,0){50}}
\put(50,30){\vector(0,1){50}}
\put(50,30){\vector(-1,-2){10}}
\thicklines
\put(50,30){\vector(0,1){30}}
\put(50,30){\vector(1,1){30}}
\put(50,30){\vector(-1,1){30}}
\put(53,80){$z$}
\put(53,60){${\vec {\bf h}}$}
\put(80,55){$\sigma_A$}
\put(15,55){$\sigma_B$}
\put(53,40){$\phi$}
\put(45,10){$x$}
\put(100,25){$y$}
\put(60,0){Fig.~1}
\end{picture}} at 0 0
\circulararc 90 degrees from -19 -15 center at -24 -20
\endpicture

\end{document}